\pdfoutput=1
\documentclass[pra,floatfix,twocolumn,superscriptaddress]{revtex4-1}
\usepackage[final]{graphicx}
\usepackage{times,bbm,amsmath,amssymb}
\usepackage{epsfig,color}
\usepackage{hyperref}
\usepackage{float,siunitx}
\usepackage[caption = false]{subfig}
\usepackage[greek,english]{babel}
\usepackage{thumbpdf,enumerate}
\usepackage{booktabs}
\usepackage{sidecap}
\usepackage[scaled=.8]{couriers}
\usepackage{pstricks}
\usepackage{multirow}
\usepackage{placeins}
\usepackage{pst-grad}
\usepackage{epigraph}
\usepackage{longtable}
\usepackage{booktabs}
\usepackage{gensymb}
\newcommand{\avW}{{\cal W}}

\usepackage{pifont}

\newcommand{\ket}[1]{\vert#1\rangle}
\newcommand{\bra}[1]{\langle#1\vert}

\begin{document}

\title{Experimental entanglement-enhanced work extraction based on a Maxwell's demon}

\author{Mario A. Ciampini}\email{marioarnolfo.ciampini@uniroma1.it}
\affiliation{Dipartimento di Fisica, Sapienza Universit\`a di Roma, P.le Aldo Moro 5, 00185, Rome, Italy}

\author{Luca Mancino}
\affiliation{Dipartimento di Fisica, Sapienza Universit\`a di Roma, P.le Aldo Moro 5, 00185, Rome, Italy}

\author{Adeline Orieux}
\affiliation{Dipartimento di Fisica, Sapienza Universit\`a di Roma, P.le Aldo Moro 5, 00185, Rome, Italy}
\affiliation{LTCI, CNRS, T\'el\'ecom ParisTech, Universit\'e Paris-Saclay, 75013, Paris, France}

\author{Caterina Vigliar}
\affiliation{Dipartimento di Fisica, Sapienza Universit\`a di Roma, P.le Aldo Moro 5, 00185, Rome, Italy}

\author{Paolo Mataloni}
\affiliation{Dipartimento di Fisica, Sapienza Universit\`a di Roma, P.le Aldo Moro 5, 00185, Rome, Italy}

\author{Mauro Paternostro} 
\affiliation{Centre for Theoretical Atomic, Molecular and Optical Physics, School of Mathematics and Physics, Queen's University Belfast, Belfast BT7 1NN, United Kingdom}

\author{Marco Barbieri} 
\affiliation{Dipartimento di Scienze, Universit\`a degli Studi Roma Tre, Via della Vasca Navale 84, 00146, Rome, Italy}

\begin{abstract}
The relation between the theory of entanglement and thermodynamics is very tight: a thermodynamic theory of quantum entanglement, as well as the establishment of rigorous formal connections between the laws of thermodynamics and the phenomenology of entanglement  are currently open areas of investigation. In this quest, an interesting problem is embodied by the role played by entanglement in processes of work extraction from a working medium embodied by quantum information carriers. In this work, we experimentally address the question \emph{``Is there any intrinsic advantage for work extraction given by the use of an entangled working medium?"}. By addressing work-extraction protocols based on a mechanism intimately linked to the paradigm of Maxwell's daemon, and implementing suitably designed multi-photon optical interferometers, we demonstrate experimentally the intrinsic advantages for such tasks provided by bipartite and genuine multipartite entanglement. We highlight the unique nature of such tests by comparing their performance to standard tests for the inseparability of multi-photon state resources. Our work contributes strongly to the ongoing efforts in establishing photonic systems as a platform for experiments for information thermodynamics. 
\end{abstract}

\maketitle

Thermodynamics is one of the pillars of physical, chemical and biological sciences~\cite{zemansky}. It predicts and explains the occurrence and efficiency of complex chemical reactions and biological processes. In physics and engineering, the conduction of heat across a medium, the concept of the arrow of time~\cite{eddington,campisi}, and the efficiency of motors are formulated in thermodynamic terms~\cite{sinitsyn}. In information theory, the {\it mantra} that ``information is physical" simply entails that, by being encoded in physical supports, information must obey the laws of thermodynamics~\cite{landauerPT}. Even more, the tightness of the link between information and thermodynamics can be appreciated from the thermodynamic interpretation of the landmark embodied by Landauer�s erasure principle~\cite{landauer}, Jaynes principle of maximum entropy~\cite{jaynes}, or the information theoretical exorcism of Maxwell's demon operated by Charlie Bennett~\cite{bennett}. 

More recently, it has been realized that thermodynamics can be used effectively to provide a new way of assessing and exploiting quantum dynamics, build super-efficient nano- and micro-engines that take advantage of quantum coherence, and assess information theory from the perspective of thermodynamic costs~\cite{goold}. 

In this context, the identification of the specific features of quantum systems that might influence their thermodynamic performance is still an open and controversial challenge. Quantum coherences are allegedly responsible for the possibility to extract work from a single heat bath~\cite{scully}. In the weak-driving quantum regime, both discrete and continuous-time heat engine exhibit non-classical signatures enabling enhanced power outputs with respect to the relevant classical (i.e. stochastic) version of such machines~\cite{uzdin}. 

On the other hand, perfectly reversible work extraction from an ensemble of non-interacting quantum information carriers appears to be possible without the need for creating any entanglement~\cite{hovhannisyan}. While entangling quantum operations allow for the retrieval of the same amount of work as in the classical case with more economic protocols, such advantage appear to fade away in the case of a large number of systems~\cite{perarnau}. 



Within such an interesting quest for the origin of ``quantumness" in quantum thermodynamics, a core achievement is given by the design of protocols allowing for an enhanced work extraction from a heat bath when using an entangled working medium rather than a separable one~\cite{maruyama,viguie}. In turn, such a result embodies a thermodynamics-inspired witness of the presence of (multipartite) entanglement in the working medium, thus contributing explicitly to the formulation of thermodynamic separability criteria. The effectiveness of such approaches to inseparability in the case of physical working media prepared in quantum laboratories has not yet been addressed due to the scarcity of experiments addressing the quantum thermodynamic framework, of which we only have a handful to date~\cite{experiments}.


In this article we bridge this gap by investigating experimentally the quantum enhanced work extraction achieved using non-ideal entangled two- and three-qubit quantum working media. We demonstrate the attainability of an experimental bound on the maximal extractable work that signals the presence of non-classical correlations. Our experiment provides evidence that non-classical correlations empowering work extraction bear differences to those responsible for the violation of Bell-type inequalities. We believe that, through so far unexplored theoretical tools, our results pioneer novel experimental methods towards the understanding of how non-classical correlations can effect the efficiency of quantum thermodynamic processes, at equal footing with thermodynamic potentials. This is a topical problem of both foundational and pragmatic relevance for the development of future engines operating at the nanoscale and in the fully quantum regime.

The remainder of this work is organised as follows: In Sec.~\ref{2} we illustrate both the theoretical scheme and the experimental implementation for entanglement-enhanced work extraction from a two-qubit photonics working medium. Sec.~\ref{3} addresses the extension to the first non-trivial multipartite scenario in this context, namely a three-photon resource prepared in either a Greenberger-Horne-Zeilinger (GHZ) state, or a $W$ one. Finally, in Sec.~\ref{final}, while drawing the salient conclusions of our investigation, we present the possible extensions of our study to other information theoretical investigations linked to the relation between (quantum) information and thermodynamics. 

\section{Two-qubit work extraction protocol} 
\label{2}

When introducing his paradigm, Maxwell had in mind Greek daimones ({\greektext da\'imones}), i.e. powerful entities bridging the human and the divine. We shall thus formulate our approach in the language of a general game involving two daemons, Aletheia ({\greektext Al\'hjeia}) and Bia ({\greektext B\'ia}), each attempting at extracting work from a two-qubit system. 

They share an ensemble of identically prepared pairs of qubits, each pair being prepared in a bipartite state $\rho$. Moreover, they agree on the choice of two sets of measurement operators (one per daemon), that densely cover a great circle on the Bloch sphere of a single qubit. For half of the ensemble, Aletheia should choose from the set $\{\hat A_\theta\}$ the projection operator to use in order to measure her qubit. In our notation, $\theta{=}0$ corresponds to the $z$ axes. For the remaining half of the ensemble, Bia shall pick her measurement operator from the set $\{\hat B_{\theta'}\}$. In these sets, $\theta$ and $\theta'$ stand for the angular position, over a great circle of their respective single-qubit Bloch sphere, of the measurements performed by the daemons. In a given run of the game, Aletheia measures operator $\hat A_\theta$ and communicates the corresponding outcome $A_\theta$ to Bia, allowing to reduce the entropy associated to her measurement result from the value $H(B_{\theta'})$ to the conditional value $H(B_{\theta'}|A_\theta)$~\cite{maruyama}. Having in mind Szilard's version of a Maxwell's daemon-inspired scenario, it is straightforward to see that the amount of work that can be extracted from a system about which we only have partial information is proportional to $1-H(a)$. Here $H(a)$ is the Shannon entropy associated with the binary variable $a$ that embodies the degree of freedom that was measured to gather information on the system. Consequently, the work Bia can extract following Aletheia's measurement and the communication of her result  is $1{-}H(B_{\theta'}|A_\theta)$. An analogous reasoning can be carried out when exchanging the role of the daemons, concluding that, for a given setting, we find that the extractable work resulting from one run of the game is 
\begin{equation}
\begin{aligned}
\label{lavoro}
W_\rho (A_\theta,B_{\theta'})&=1-\frac{1}{2}[H(A_\theta|B_{\theta'})+H(B_{\theta'}|A_\theta)]\\
&=1-H(A_\theta,B_{\theta'})+\frac12[H(A_\theta)+H(B_{\theta'})],
\end{aligned}
\end{equation}
where $H(A_\theta,B_{\theta'})$ is the joint entropy of variables $A_\theta$ and $B_{\theta'}$. With the choices made above, $W_\rho (A_\theta,B_{\theta'})$ is measured in bits. For a sufficiently dense covering of the great circle, and for ${\theta'}=\theta$, the average extractable work achieved through the game, maximised over all the possible great circles, is
\begin{equation}
\label{lavoromed}
{\cal W}(\rho)= \frac{1}{2\pi}\max_{\phi}\int_0^{2\pi} W_\rho (A_\theta,B_\theta)d\theta.
\end{equation}
where $\phi$ is the azimuthal angle for the great circles being considered.  In Ref.~\cite{viguie} it was shown that any separable state $\rho$ is such that 
\begin{equation}
\label{dis}
\avW(\rho) \leq \avW_{f}=0.443\,\text{bits}.
\end{equation}
where $\avW_f$ is the maximum average work evaluated for any pure factored state. Such upper bound represents the maximum work extractable, through such a game, from classically correlated states. Therefore, any surplus of extractable work above $\avW_f$ arises in virtue of entanglement shared by the daemons. Eq.~\eqref{dis} can then be used as a thermodynamically rooted entropic witness for inseparability. 
While entropic inequalities have been previously introduced and tested~\cite{entropic}, these are rooted in purely informational considerations, while here we have a direct connection with thermodynamics.
\begin{figure}[t!]
\includegraphics[width=\columnwidth]{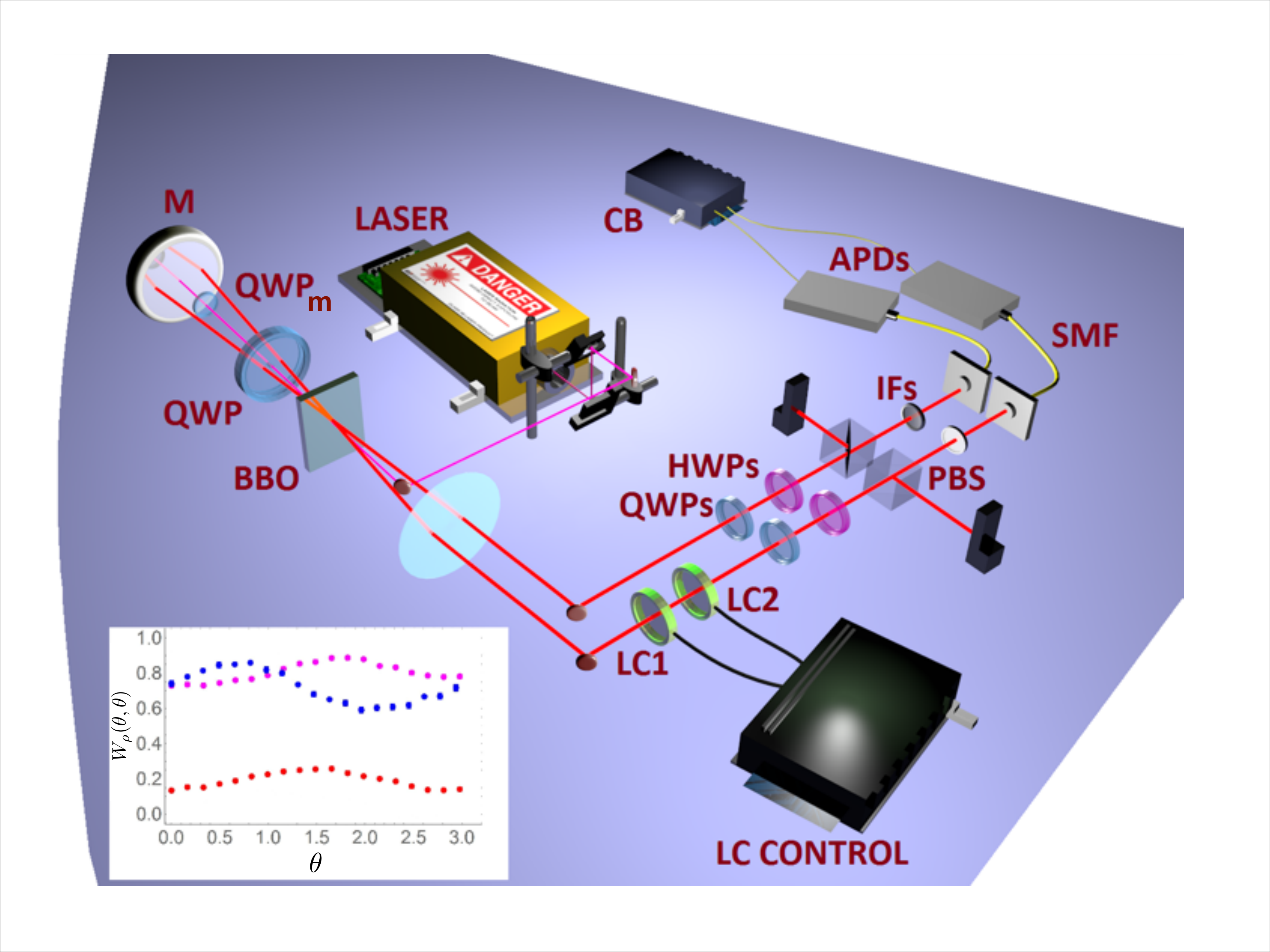}
\caption{Experimental setup for the extraction of work from inseparable bipartite state: The entangled photon source uses a 1.5mm thick $\beta$ Barium-Borate (BBO) crystal pumped with 100 mW of laser at 355 nm, in conjunction with a spherical mirror (M), and delivers approximately 200 coincidences/s through 5 nm (FWHM) interference filters (IF). Details on the source can be found in Ref.~\cite{dinepi}. We encode the logical states of each qubit in the horizontal and vertical polarization states $\ket{H}$ and $\ket{V}$ of each photon. Using quantum state tomography, we estimate a fidelity $F=0.961\pm0.007$ of the entangled resource with the maximally entangled state $\frac{1}{\sqrt{2}}\left(\ket{HH}+\ket{VV}\right)$. The corresponding value of tangle is $T=0.911\pm0.008$. The relative weight ($\cos\varphi$) of the two polarization contributions in such state can be tuned by the quarter waveplate (${\rm QWP}_{\rm m}$) in the source. One of the photons passes through a depolarizing channel, consisting of two liquid crystals (LCs) and the associated control electronics, which selects the value of $\mu$. Finally, polarization measurements are performed at a polarisation-analysis module consisting of  a QWP, a half waveplate (HWP), a polarising beam splitter (PBS), and an avalanche photodiode (APD) per mode. We also show the single-mode fibers (SMF) used to convey the photonics signal to the polarization analysis module. Inset: measured $W_\rho(\theta,\theta)$ as a function of the measurement angle $\theta$ for the Eq.~\eqref{state} with $\cos\varphi=0.85$, and $\mu=0.98$ (blue); $\cos\varphi=0.62$, and $\mu=0.98$ (purple); $\cos\varphi=0.62$, and $\mu=0.51$ (red).}
\label{fig:exp_setup}
\end{figure}


We employ the thermodynamic separability criterion in Eq.~\eqref{dis} for the characterisation of photonic entanglement. Polarization-entangled pairs of qubits, ideally prepared in the state $|\Phi \rangle{=}{\cos \varphi |HH\rangle + \sin \varphi |VV\rangle}$, are produced by using the double-pass source in~\cite{dinepi} (see Fig.~\ref{fig:exp_setup}). Here, $\ket{H}$ ($\ket{V}$) indicates a photon in the horizontal (vertical) polarization state. 
Further, we explore how the thermodynamic bound evolves through a depolarising channel degrading the initial state to the mixture
\begin{equation}
\label{state}
\rho=\mu |\Phi\rangle \langle \Phi| + \frac{1-\mu}{4} \mathbb{I}.
\end{equation}
Here, the parameter $1{-}\mu$ quantifies the strength of the channel. This has been implemented by using two liquid crystals, with their axes at $0{\si{\degree}}$ and $45{\si{\degree}}$, onto the path of one photon. The two crystals are tuned in such a way that they perform a rapid succession of Pauli operators  $\hat{\sigma}_x$, $\hat{\sigma}_y$, $\hat{\sigma}_z$, thus generating white noise on the state~\cite{channels}.

\begin{figure*}[t!]
{\bf (a)}\hskip8cm{\bf (b)}
{\includegraphics[width = \columnwidth]{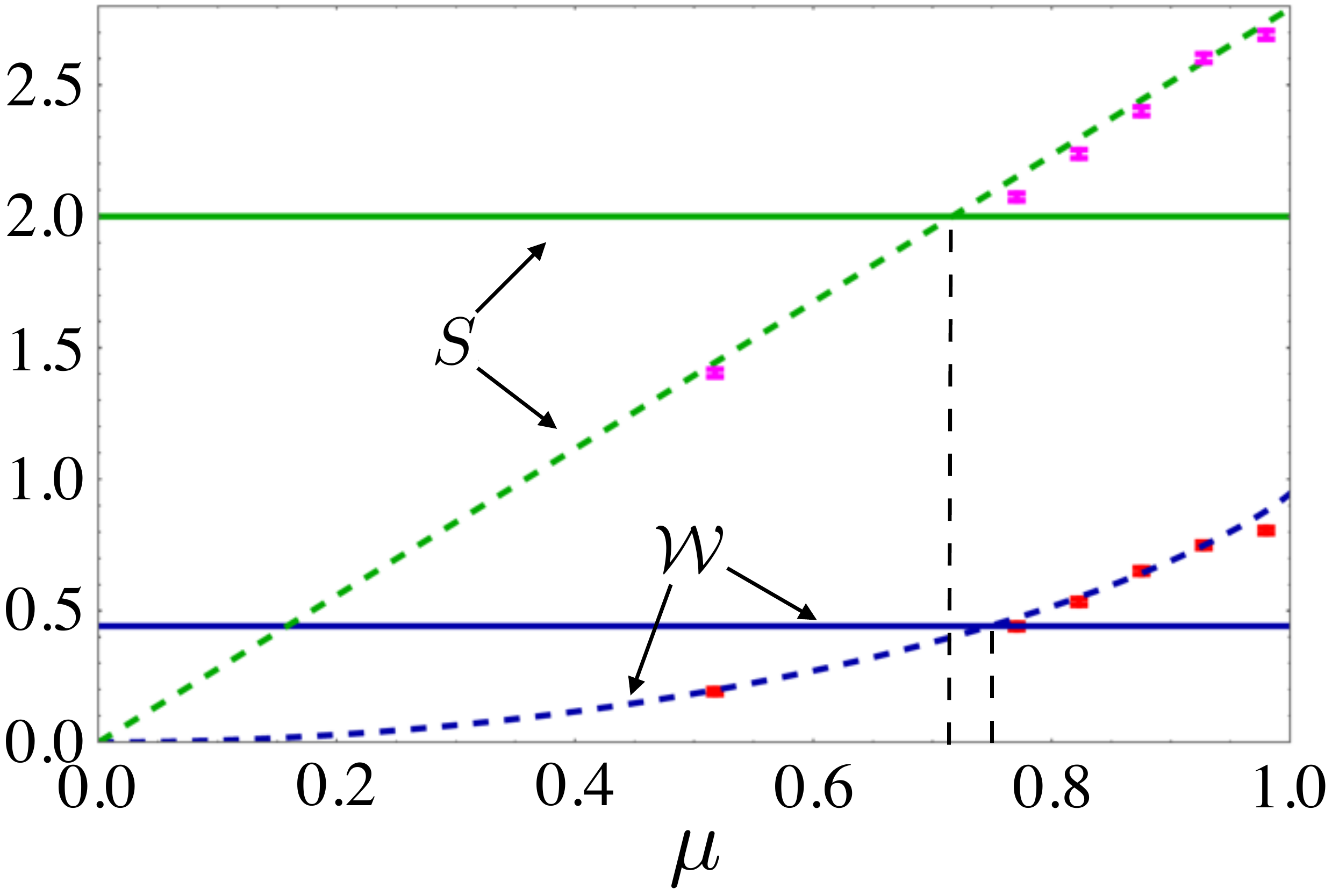}}~~{\includegraphics[width = \columnwidth]{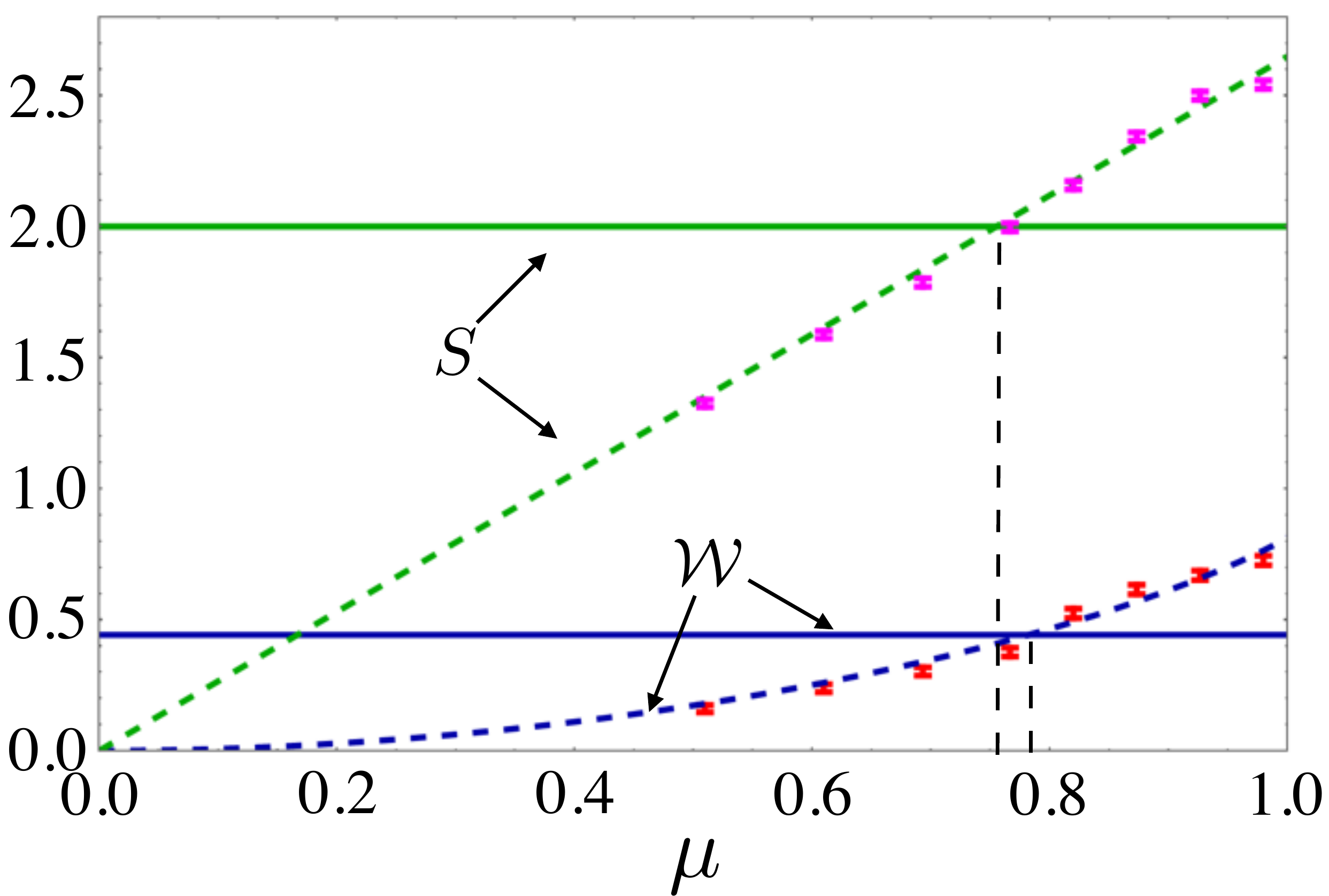}} 
\caption{Experimental results for two different resource states corresponding to $\cos\varphi=0.62$ [panel {\bf (a)}], and $\cos\varphi=0.85$ [panel {\bf (b)}] for different strengths of the depolarising channels $\mu$. Red dots: experimental points for $\avW$ calculated from the experimental count rates measured in 30s. Magenta dots: experimental points corresponding to the values taken by the Bell parameter, evaluated from the fully reconstructed state tomographies of the resource $\rho_{\rm exp}$. The value of $\cos\varphi$ is evaluated by direct inspection of the coincidence rates, averaged over all the experimental states. The error bars take into account the Poissonian statistics of the measured rates and derived from direct error propagation for $\avW$ and Monte Carlo simulation for $S$. The green and blue dashed curves represent theoretical predictions based on Eq.~\eqref{state}, while the solid straight lines denote the entanglement thresholds for the work extraction and the Bell function. The dashed vertical lines identify the values of $\mu$ for which $\avW$ and $S$ cross their respective bounds.}
\label{fig:exp_data}
\end{figure*}

In the inset of Fig.~\ref{fig:exp_setup}, we show typical experimental curves for the work $W_\rho(\theta,\theta)$ calculated according to Eq.~\eqref{lavoro}, as we inspect the set of operator pairs used by Aletheia and Bia, labelled by the angle $\theta$ on the great circle of the Bloch sphere corresponding to the linear polarizations (i.e. the equator of the single-qubit Bloch sphere). This is the set of directions allowing for optimal work extraction. The observed oscillations are due to the polarization unbalance, while the average level is affected by the purity of the state. For each choice of $\theta$, we have calculated the Shannon entropy associated to  single-particle operators, as well as the joint entropy associated to the joint operator. We have verified by numerical simulations that our covering with 19 different directions is sufficiently dense for the continuous approximation to hold. In addition, we have fully characterised our states with quantum state tomography, from which the maximal value of the Bell operator $S$ has been extracted for a comparison with a standard entanglement witness. 

In order to perform the average over the projections performed by the daemons, we removed the quarter waveplates (QWPs) from the polarization analysis-module shown in Fig.~\ref{fig:exp_setup}, and used the half waveplates (HWPs) to rotate the state along the correct big circle. The effect of the polarization beam splitter (PBS) is that of performing the projection $\vert H \rangle \langle H \vert$. As we used just two avalanche photo-diodes (APDs) after the PBSs, we had to implement four different types of measurements for each of the 19 choices of angular direction on the equator of the single-qubit Bloch sphere. In particular, we had to consider the four sets of directions $(\theta_A,{\theta'}_B)=(\theta,\theta), (\theta,\theta+45\si{\degree}), (\theta+45\si{\degree},\theta), (\theta+45\si{\degree},\theta+45\si{\degree})$, where the first (second) angular direction is for Aletheia's (Bia's) measurements choice. The corresponding detected coincidences at the APDs are labelled as $N_{\theta_A,{\theta'}_B}$. A full circulation of the Bloch-sphere equator is then achieved by rotating the HWPs from $0\si{\degree}$ to $45\si{\degree}$ in 19 steps. This implies that a total of 76 measurements were needed in order to acquire the  necessary information to perform the evaluation of the work extraction performance. Upon suitable normalization of the detected coincidence counts, we have the following set of probabilities 
\begin{align}
p_{AB}(\theta)&={N_{\theta,\theta+45\si{\degree}}}/D(\theta), \\
p_{A}(\theta)&=({N_{\theta, \theta}+N_{\theta+45\si{\degree},\theta}})/D(\theta),\\
p_{B}(\theta)&=({N_{\theta, \theta}+N_{\theta,\theta+45\si{\degree}}})/D(\theta)
\end{align}
with $D(\theta)=N_{\theta,\theta}+N_{\theta,\theta+45\si{\degree}}+N_{\theta+45\si{\degree}, \theta}+N_{\theta+45\si{\degree},\theta+45\si{\degree}}$. 
These are instrumental to the evaluation of the Shannon entropies needed to calculate $W_\rho(\theta,\theta)$ and, in turn, $\avW$.

\begin{figure*}[t!]
{\bf (a)}\hskip9cm{\bf (b)}
\includegraphics[width =\columnwidth]{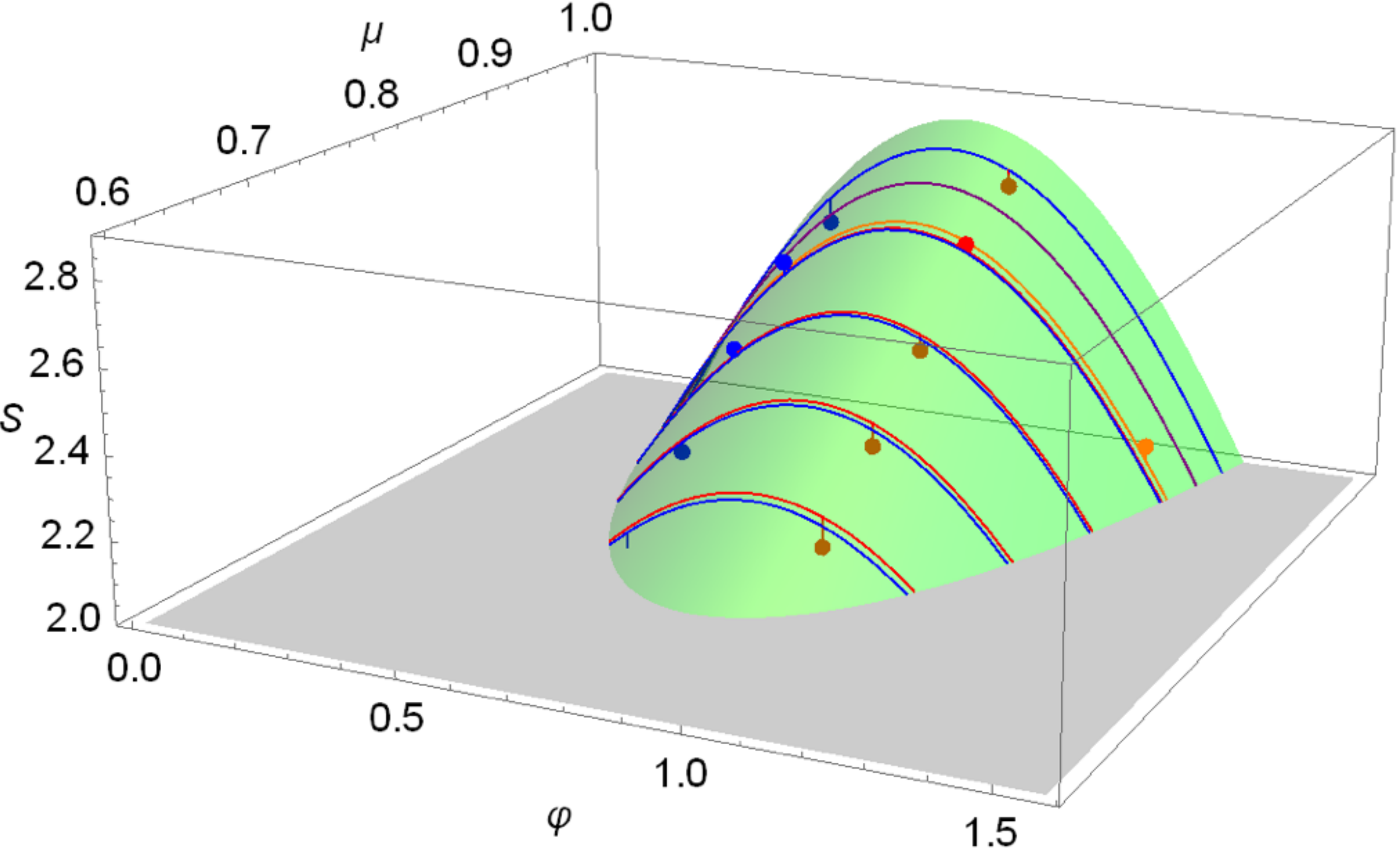}\includegraphics[width = 1.1\columnwidth]{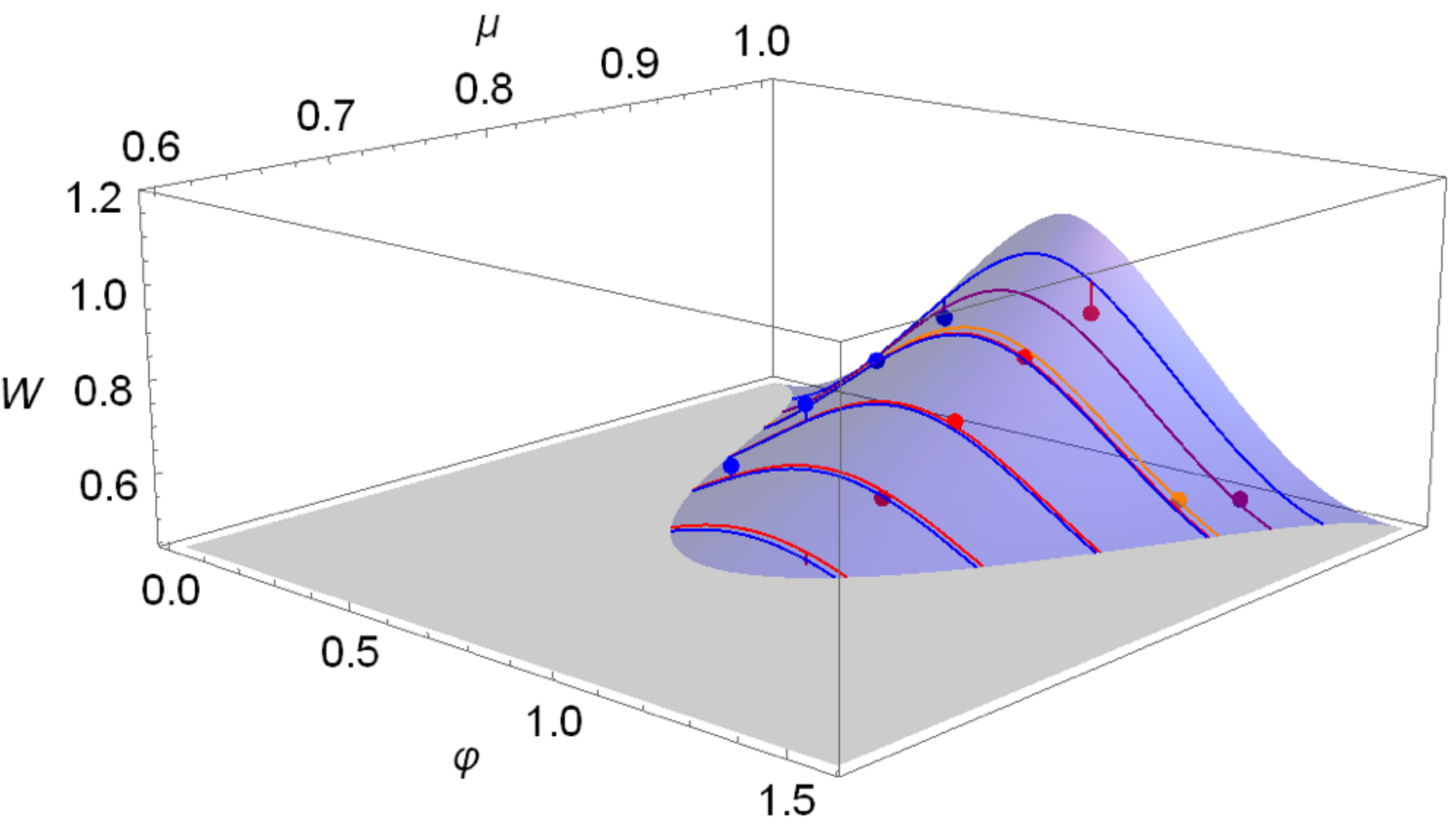}
\caption{Comparison between theoretical predictions and experimental data for the Bell function [panel {\bf (a)}] and the extractable work [panel {\bf (b)}] against the strength $\mu$ of the channel and the state unbalance $\varphi$ (measured in degrees). The semi-transparent surfaces show the behavior of $S$ and $\avW$, while the dots represent the experimental points. Only the parts of the theoretical surfaces exceeding the local realistic bound [panel {\bf (a)}] and the threshold $\avW_f$ [panel {\bf (b)}] are displayed. Moreover, in order to highlight  the relative position of the experimental points with respect to the theoretical surfaces, we show the curves on the latter associated with the same value of $\mu$.}
\label{fig:3D_graphs}
\end{figure*}

We measured the extractable thermodynamic work for several sets of parameters $\varphi$ and $\mu$ in the resource state, which were estimated by reconstructing the density matrix of the experimentally engineered state $\rho_{\rm exp}$ via complete quantum state tomography for each position of the QWP in the source (setting $\varphi$) and every configuration of the LCs (fixing $\mu$). 
Extractable work has been calculated by measuring Shannon entropy spanned over the whole big circle of the Bloch Sphere.  Example of measured coincidence counts are $N_{0,0}=3578$, $N_{0,45\si{\degree}}=58$, $N_{45\si{\degree},0}=173$, $N_{45\si{\degree},45\si{\degree}}=4328$, achieved in 30s of measurement.

In order to highlight the inherently different nature of the test implemented here and a standard Bell test, we evaluate the Bell function as 
\begin{equation}
S={\rm Tr}[\rho(\hat O_1-\hat O_2+\hat O_3+\hat O_4)]
\end{equation}
with $\hat O_1=\sigma_x\otimes\sigma_x(\alpha)$, $\hat O_2=\sigma_x\otimes\sigma_z(\alpha)$, $\hat O_3=\sigma_z\otimes\sigma_x(\alpha)$, and $\hat O_4=\sigma_z\otimes\sigma_z(\alpha)$. Here, $\sigma_k(\alpha)=\hat R(\alpha)\sigma_k\hat R^\dag(\alpha)$ with $\hat R(\alpha)=\cos\alpha\openone+i\sin\alpha\sigma_y$ and $\sigma_k$ is the $k=x,y,z$ Pauli matrix. Local realistic theories bound such function as $|S|\le2$, while quantum mechanically $|S|\le2\sqrt2$. Using Eq.~\eqref{state}, we find 
\begin{equation}
S=\mu[1+\sin(2\varphi)][\cos(2\alpha)+\sin(2\alpha)],
\end{equation}
which we have used to estimate the value of the Bell function using the experimental density matrix $\rho_{\rm exp}$ for each set value of $\mu$ and $\varphi$.

The results of our experiment are summarised in Fig. 2, where we show the measured extractable work and the Bell function for two resource states \eqref{state} with different values of $\varphi$. The comparison with Bell's test reveals how both quantities capture the degradation resulting from the depolarisation channel, and that they are similarly robust against the bias between the $\ket{HH}$ and $\ket{VV}$ contributions. A complete analysis shows that the relation between the thermodynamic criterion and Bell's inequalities is not trivial. In fact, we were able to find values of $\mu$ and $\varphi$ for which it is possible to violate just one of the two criteria, and vice versa. We have considered an experimental state with large bias ($\cos\varphi=0.25\pm0.03$) and purity ($\mu=0.97$) that offers no violation of the local realistic bound ($S=1.977\pm0.009$). However, the corresponding extractable work is $\avW=0.491\pm0.021$, which exceeds the separability threshold $\avW_{f}$ by 2 standard deviations. This implies that the use of entanglement for powering a thermal machine is qualitatively different from that necessary for communication tasks, such as measurement-independent key distribution. Fig.~\ref{fig:3D_graphs} shows a comprehensive comparison between the theoretical predictions gathered by using Eq.~\eqref{state} and the experimental data, revealing the great agreement between the latter and the former throughout the whole range of parameters assessed in our experimental investigation.

\begin{table}[b!]
    \label{tab:table1}
\begin{center}
\begin{tabular}{ccccc} 
\hline
\hline
$\mu$ & $\varphi$ & $\avW$ & $S$ & $T$ \\
\hline
\hline
0.51 & 0.91(4) & 0.193(5) & 1.404(7) & 0.04(1) \\
0.77 & 0.91(4) & 0.441(6) & {\bf 2.073(7)} & 0.34(4) \\
0.82 & 0.91(4) & {\bf 0.534(6)} & {\bf 2.237(8)} & 0.48(5) \\
0.87 & 0.91(4) & {\bf 0.652(6)} & {\bf 2.400(8)} & 0.60(5) \\
0.92 & 0.91(4) & {\bf 0.749(6)} & {\bf 2.602(8)} & 0.81(6) \\
0.98 & 0.91(4) & {\bf 0.805(6)} & {\bf 2.690(8)} & 0.91(4) \\
\hline
0.51 & 0.54(7) & 0.160(7) & 1.325(7) & 0.03(1) \\
0.61 & 0.54(7) & 0.238(8) & 1.587(7) & 0.13(2) \\
0.69 & 0.54(7) & 0.302(8) & 1.786(8) & 0.20(2) \\
0.74 & 0.54(7) & 0.376(8) & 1.997(8) & 0.34(3) \\
0.82 & 0.54(7) & {\bf 0.524(9)} & {\bf 2.157(8)} & 0.45(2) \\
0.87 & 0.54(7) & {\bf 0.615(9)} & {\bf 2.343(8)} & 0.62(3) \\
0.92 & 0.54(7) & {\bf 0.669(9)} & {\bf 2.499(8)} & 0.78(3) \\
0.98 & 0.54(7) & {\bf 0.726(9)} & {\bf 2.541(8)} & 0.82(3) \\
\hline
0.93 & 1.25(4) & {\bf 0.501(12)} & {\bf 2.131(9)} & 0.34(4) \\
\hline
0.95 & 1.32(4) & {\bf 0.491(21)} & 1.977(9) & 0.24(4) \\
\hline
\hline
\end{tabular}
\end{center}
\label{tab:exp_data}
\caption{Table of the experimental data. Here $\mu$ is the strength of the channel, while $\varphi$ determines the unbalance between the $|HH\rangle$ and $|VV\rangle$ components in the ideal resource state $\ket{\Phi}$. Both parameters have been estimated by performing state tomography of the experimental resource $\rho_{\rm exp}$; $\avW$ is the extractable work from the system. Its standard deviation $\Delta \avW$ has been obtained by considering Poissonian statistic on the coincidence counts. $S$ is the Bell function, while $T$ provides the value of tangle~\cite{horodecki}, which quantifies the degree of entanglement within the state of the system. Both these parameters (and associated uncertainties) have been estimated using state tomography. Values of $\avW$ and $S$ reported in bold violate the separability and local realistic bound, respectively.}
\end{table}

\section{Three-qubit work extraction}
\label{3}

\begin{figure}[t!]
\includegraphics[width=\columnwidth]{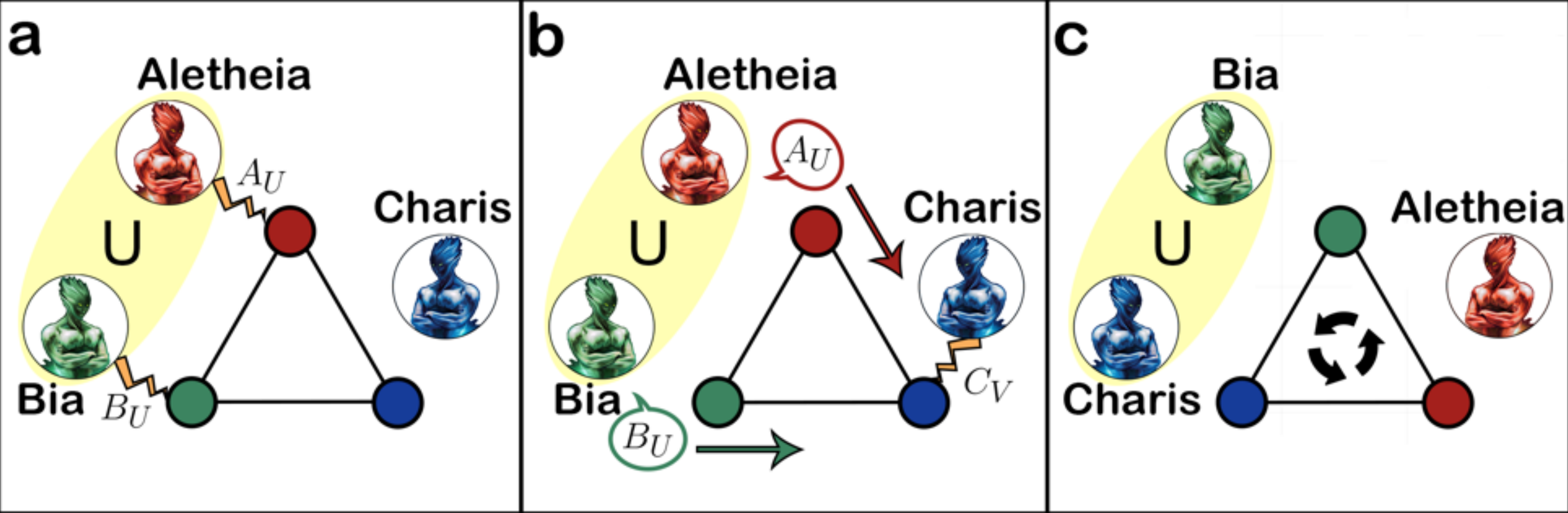}
\caption{Diagrammatic representation of the scheme for work extraction from a tripartite system. Each panel illustrates one of the steps of the protocol described in the main text. }
\label{fig:schemaTripartito}
\end{figure}

We now extend our investigation to the case of a tripartite system, which naturally present a richer entanglement-sharing structure, and embody the first non trivial instance of systems exhibiting genuine multipartite entanglement~\cite{horodecki}. In this scenario, both the figure of merit presented in Eq.~\eqref{lavoro} and the threshold for inseparability based on extractable work need to be adjusted to capture the above mentioned richness. In particular, it is well known that, for pure states, only two non-equivalent classes of genuine multipartite entanglement exist~\cite{dur}, either the GHZ or $W$ class, the situation being more complex when mixed states are considered. In our investigation, we focus on the ability of distinguishing GHZ-type entanglement from W-type one using thermodynamics bound, as opposed to state fidelity~\cite{weinfurter}. The extractable work-based inseparability criterion should thus incorporate the possibility for the resource that we address to belong to either class.

The protocol now involves a further daemon, Charis ({\greektext Q\'aris}), which is set to extract work from her subsystem, based on information provided by both Aletheia and Bia. The strategy they agreed on consists of the following steps: 
\begin{enumerate}[{\bf a}]
\item Aletheia and Bia perform a projective measurement along a common axis $u$ in the single-qubit Bloch sphere chosen among the three Pauli settings.
\item Charis receives information on the outcomes of such measurements. In light of such information, she can extract work from her system by performing a projective measurement along the suitably chosen direction $v$ on the Bloch sphere. The amount of work that can be obtained in a single run of such a tripartite game is then
\begin{equation}
W_{\rho}(A_u,B_u,C_v)=1-H(C_v|A_u,B_u),
\end{equation}
with $H(C_v|A_u,B_u)$ the Shannon entropy of variable $C_v$  conditioned on the outcomes $A_u$ and $B_u$ performed by Aletheia and Bia. 
A bound for the average extractable work, obtained by maximizing over the choice of $v$, can be established for separable states as
\begin{equation}
\begin{aligned}
\label{ghz}
\avW(\rho)=\max_v \,\frac{1}{3}\sum_{u} W_{\rho}(A_u,B_u,C_v)\leq\avW_f=\frac{1}{3}.
\end{aligned}
\end{equation}
A violation of this bound signals the presence of entanglement in the state. 



\begin{figure}[t!]
{\bf (a)}
\includegraphics[width=\columnwidth]{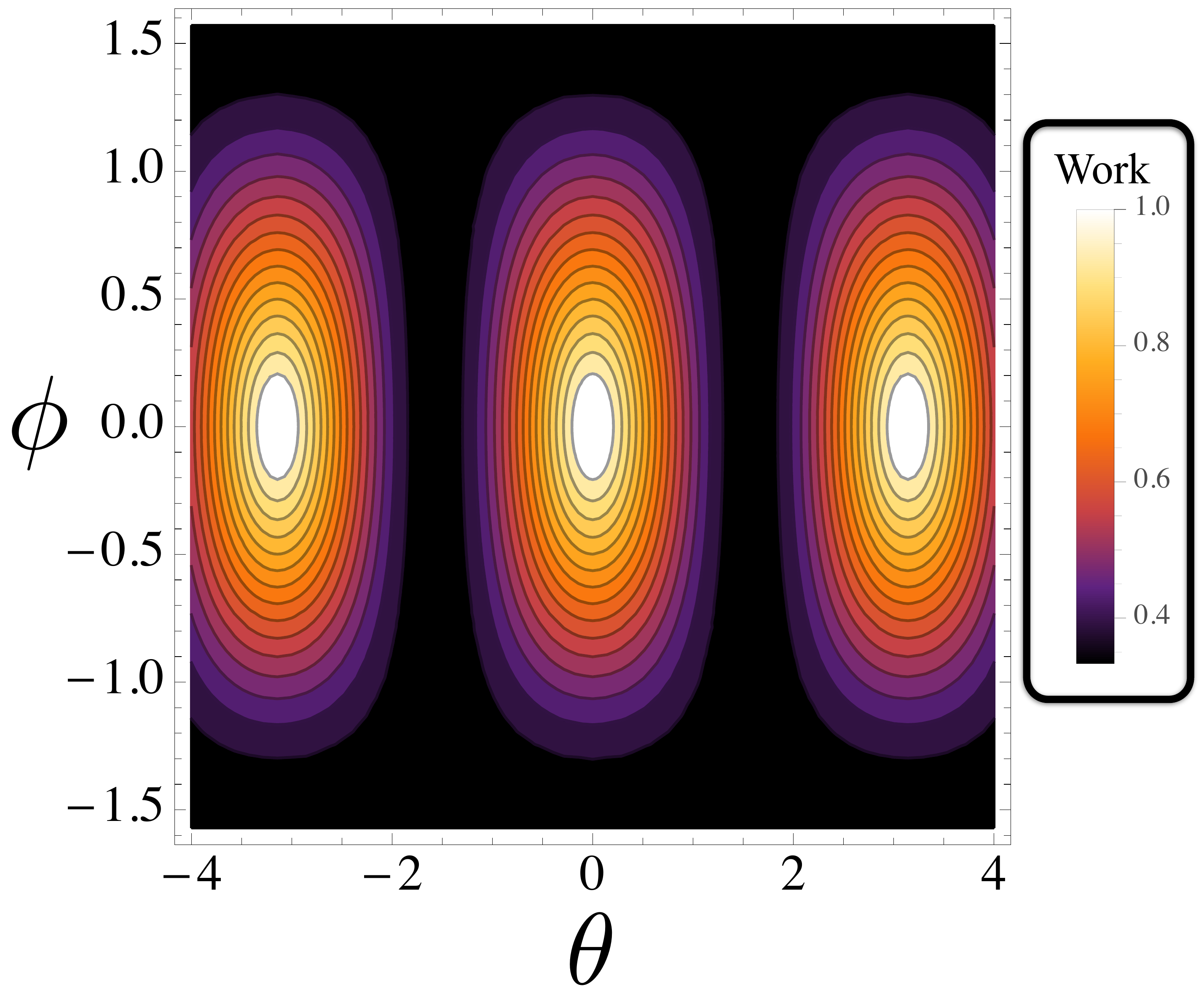}\\
{\bf (b)}
\includegraphics[width=\columnwidth]{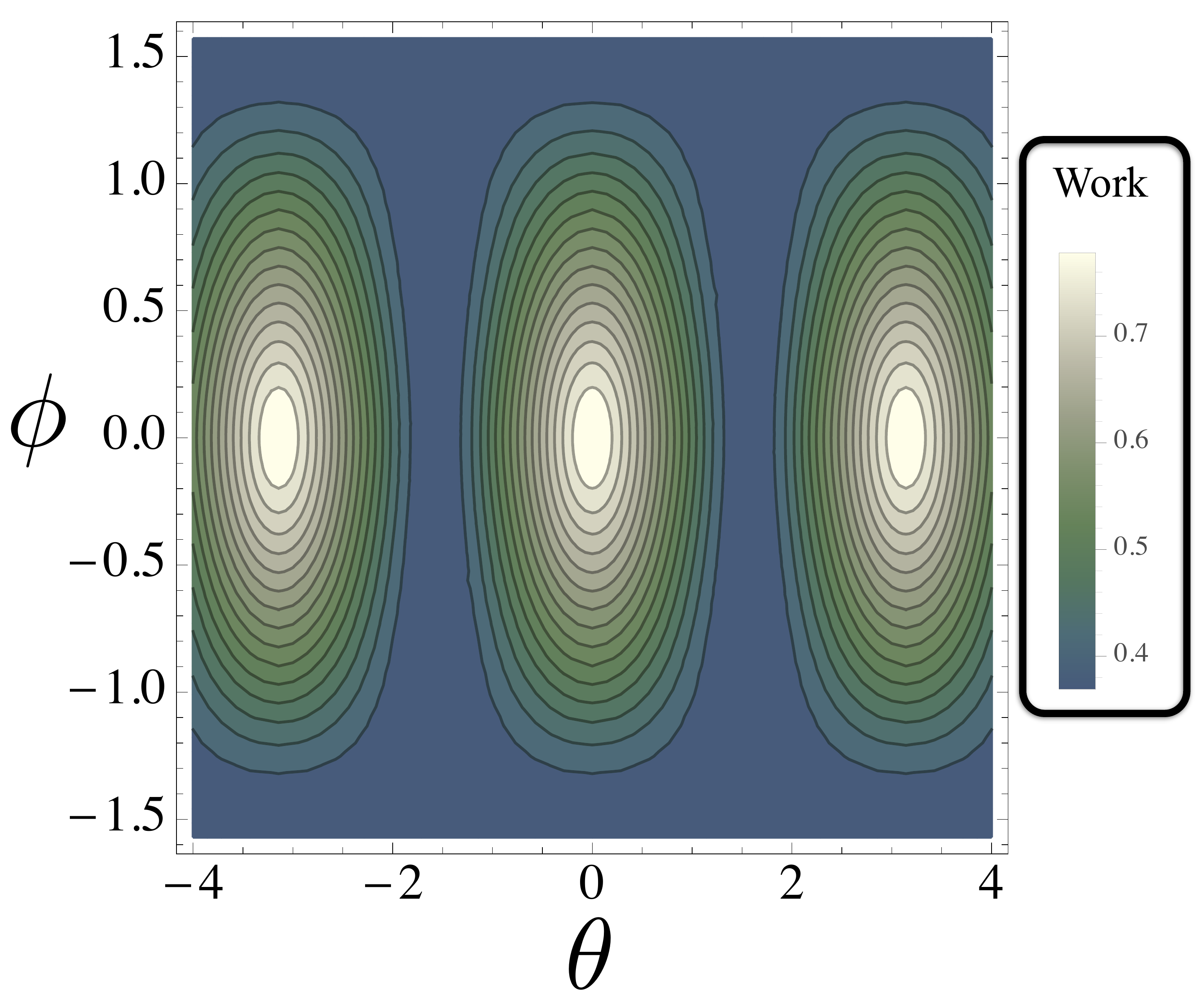}
\caption{Density plot of the theoretical work that can be extracted from pure-state tripartite resources against the angles $(\theta,\phi)$ that identify the angular direction of the Bloch sphere in which $\avW(\rho)$ can be optimized. Panel {\bf (a)} is for a pure GHZ-Cluster state, while panel {\bf (b)} shows the behavior for a $W$ state. Evidently, the maximum work achievable in panel {\bf (a)} can be larger than that resulting from using a $W$ state. }
\label{fig:simulTripartito}
\end{figure}



\begin{figure}[t!]
{\bf (a)}
\includegraphics[width=\columnwidth]{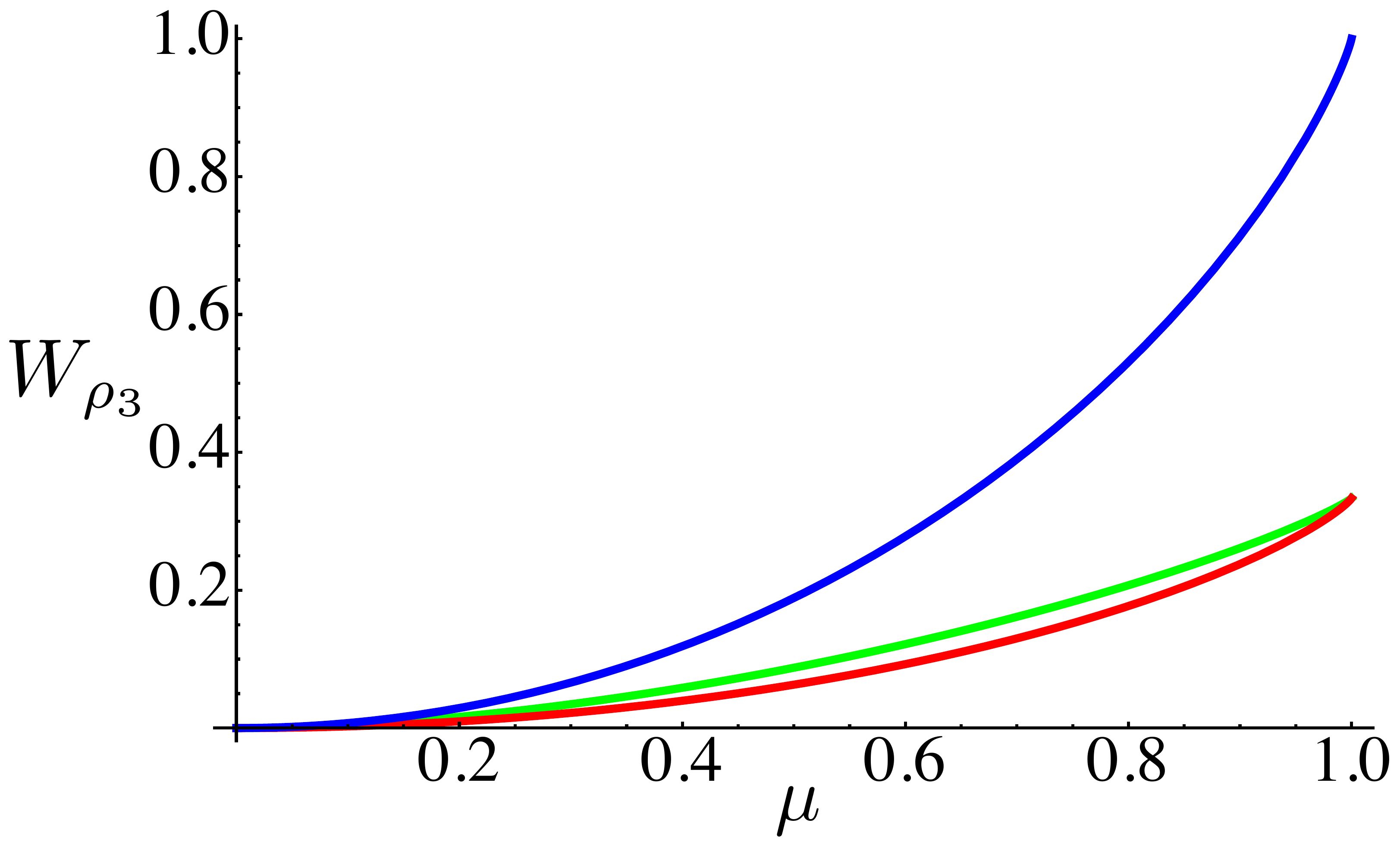}\\
{\bf (b)}
\includegraphics[width=\columnwidth]{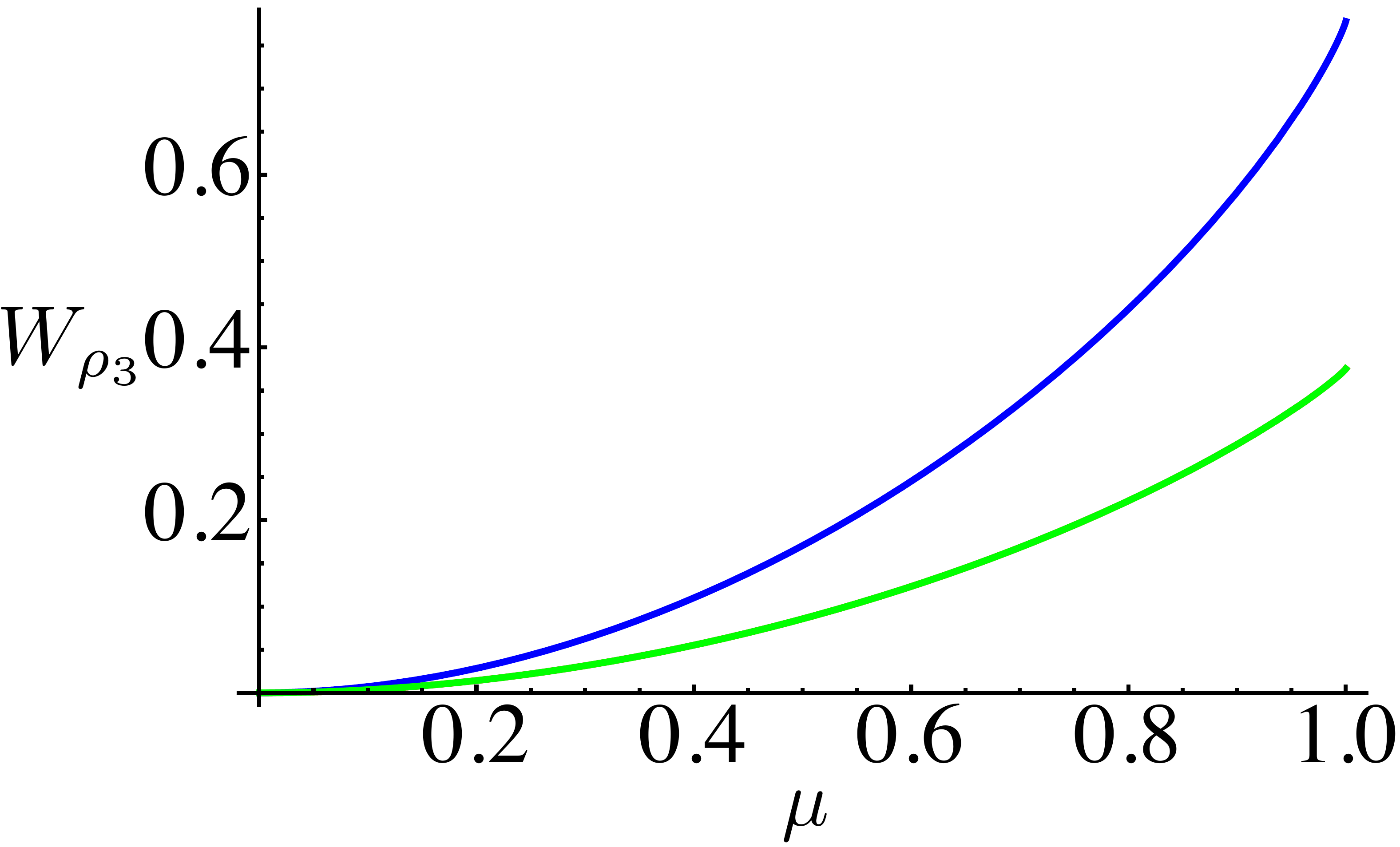}
\caption{Theoretical work that can be extracted using a resource state embodied by $\rho_{3}$ with $\ket{\Psi}$ being a GHZ-Cluster state [panel {\bf (a)}] or a $W$ state [panel {\bf (b)}], studied against $\mu$ and for the three possible choices of U such as $\hat{\sigma}_x$, $\hat{\sigma}_y$, $\hat{\sigma}_z$. In both panels the red curve is for $(\theta,\phi)=(90^{\si{\degree}},0^{\si{\degree}})$, the blue one is for $(\theta,\phi)=(0^{\si{\degree}},0^{\si{\degree}})$ and the green one for $(\theta,\phi)=(0^{\si{\degree}},90^{\si{\degree}})$. In panel {\bf (b)} the red and green curves are superimposed, implying that the work that can be extracted choosing these directions on the Bloch sphere is the same.}
\label{fig:simulWernerTripartito}
\end{figure}

\begin{figure}[b!]
{\bf (a)}
\includegraphics[width =\columnwidth]{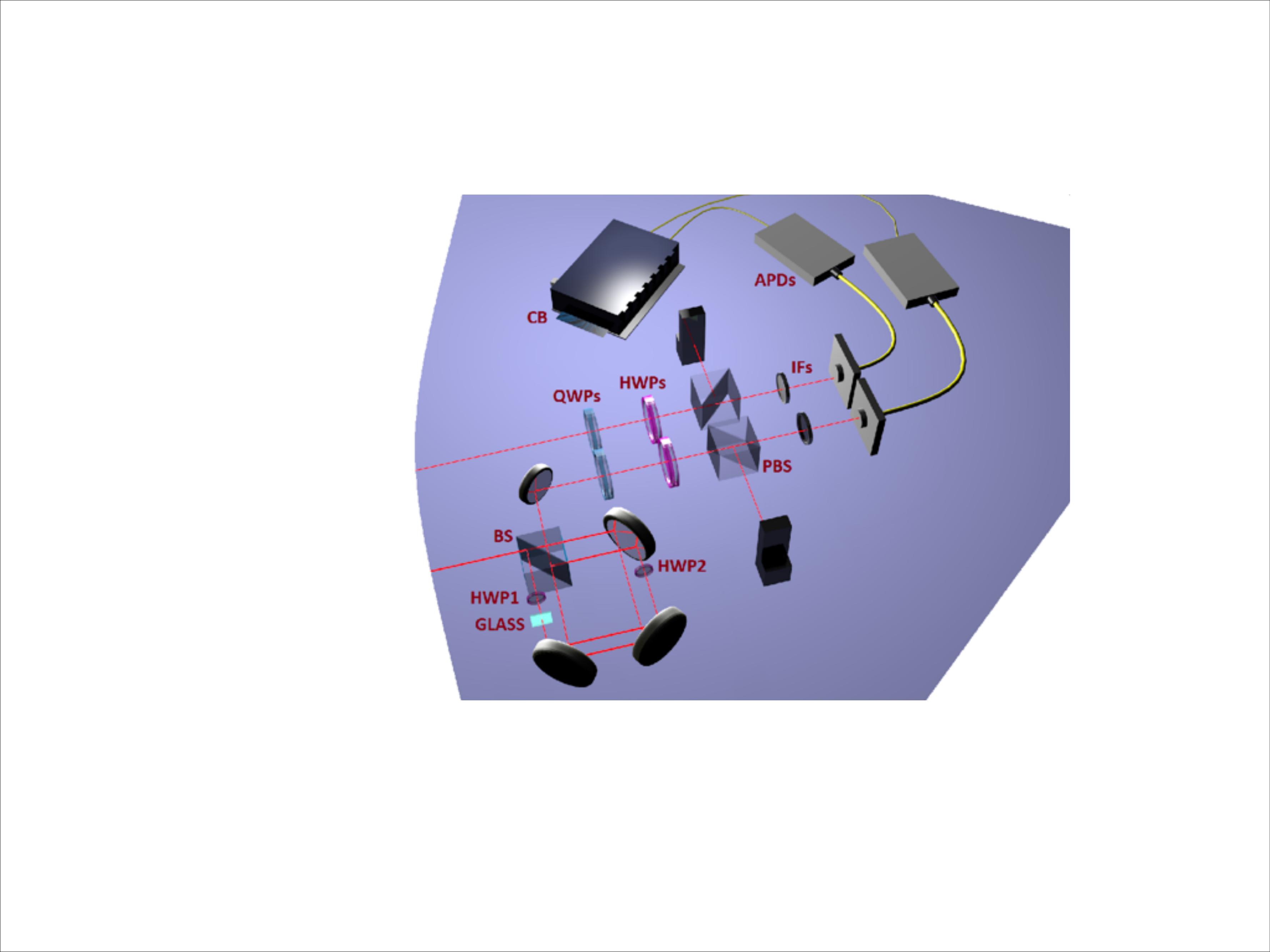}\\
{\bf (b)}
\includegraphics[width =\columnwidth]{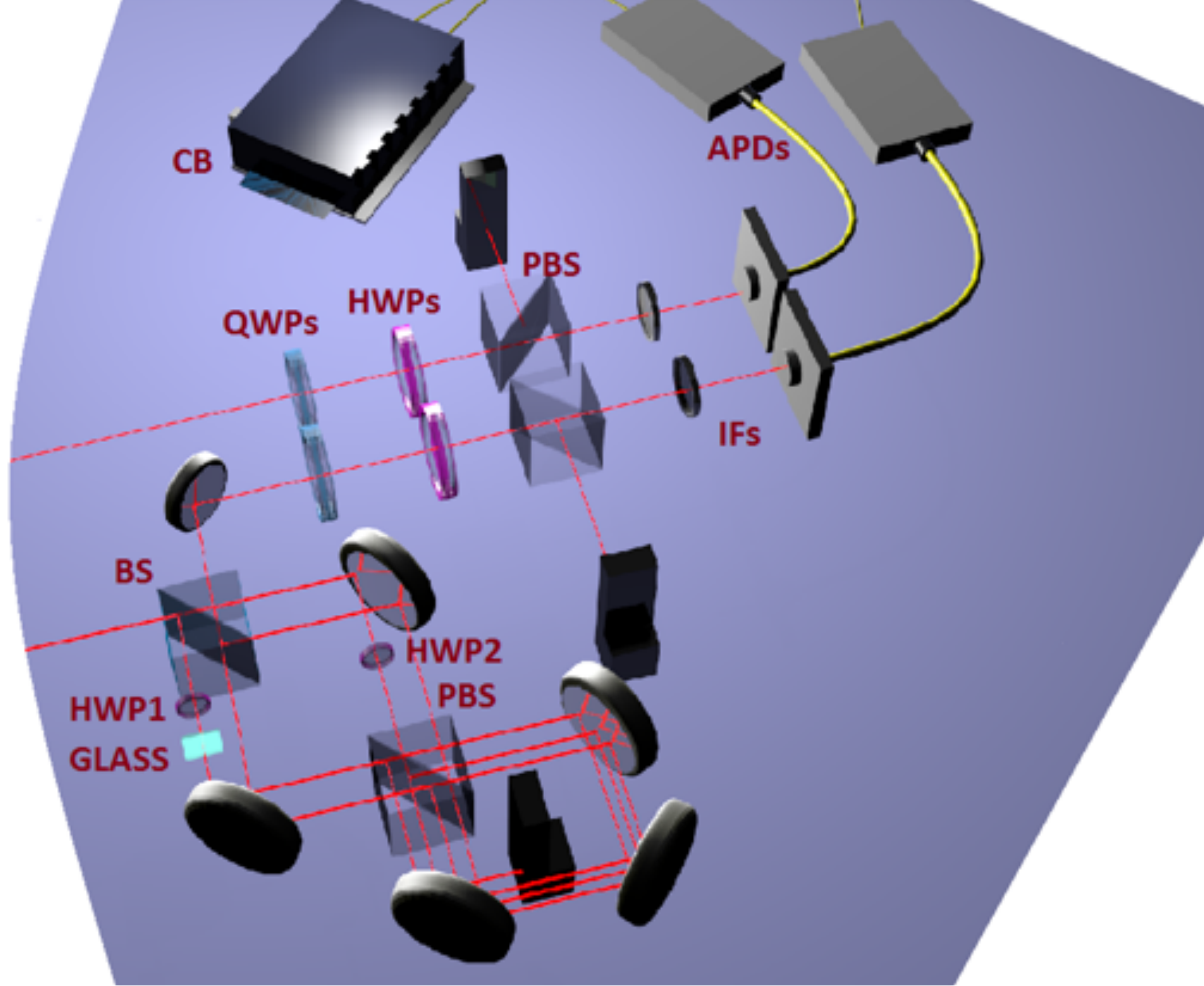}
\caption{Experimental setup for the extraction of work from a GHZ-Cluster state [panel {\bf (a)}] and a $W$-type resource [panel {\bf (b)}]. The symbols and notation used in both panels are the same as in Fig.~\ref{fig:exp_setup}.}
\label{fig:set-upGHZ}
\end{figure}

\begin{table}[b!]
  \begin{center}
    \begin{tabular}{ccc}
\hline
\hline
      $\avW^{GHZc}_1$ & $\avW^{GHZc}_2$ & $\avW^{GHZc}_3$\\
      \hline
      $0.798\pm0.006$ & $0.779\pm0.005$ & $0.795\pm0.004$\\
      $0.799\pm0.006$ & $0.304\pm0.003$ & $0.305\pm0.003$\\
      \hline
      \hline
    \end{tabular}
  \end{center}
 \caption{Extractable work in bits for the three permutations of roles in the protocol for work extraction-based inseparability evaluated using a $GHZ$-Cluster resource. The three columns are for the value of the extractable work achieved from the path-encoded qubit, the first polarization-encoded qubit, and the second polarization-encoded one, respectively. The first row refers to a tripartite entangled GHZ-Cluster state  whose fidelity with the ideal resource $(\ket{HH0}-\ket{VV0}+\ket{HV1}+\ket{VV1})/2$ is found to be $F=0.851\pm0.008$. The second row refers to the same resource where any coherence between path and polarization qubit is removed. Correspondingly, $\avW^{GHZ}_2$ and $\avW^{GHZ}_3$ are below the threshold for separability, showing that showing that in the  corresponding configuration we only have bipartite entanglement.  }
     \label{tab:table2}
\end{table}

\item In order to exclude bipartite entanglement and thus link the excess extractable work to the presence of genuine multipartite entanglement in our resource state, the protocol must be repeated by permuting the role of the three daemons. Furthermore, it can be shown that $W$-type entanglement can only provide an extractable work of at most ${7}/{9}$ bits~\cite{viguie}. Therefore, this thermodynamic witness is well suited for identifying the entanglement class to which the state belongs: any extractable work in excess of $7/9=0.778$ bits would signal the GHZ character of the resource being used.  
\end{enumerate}

In order to achieve useful benchmarks for the performance of our experiments, we have performed a thorough theoretical analysis of the protocol for both pure and mixed-state resources. In particular, we have considered the achievable value of the extractable work $\avW$ when using the resource state
\begin{equation}
\rho_3=\mu\ket{ \Psi}\bra{\Psi}+\frac{1-\mu}{8} \mathbb{I}
\end{equation}
with $\ket{\Psi}$ a tripartite pure state. We have first focused on the case of $\mu=1$ with $\ket{\Psi}$ embodied by either a tripartite GHZ-Cluster state of the form
\begin{equation}
\label{ghzcluster}
\ket{GHZc}=\frac12(\ket{000}-\ket{110}+\ket{011}+\ket{101})_{ABC}
\end{equation}
where $\{\ket{0},\ket{1}\}$ are the logical states of the three qubits, labelled as $A$, $B$, and $C$ and owned by Aletheia, Bia, and Charis respectively. This state can be shown to give rise to the same work extraction performance as a standard GHZ resource. The $W$-type resource is instead
\begin{equation}
\label{W}
\ket{W}=\frac{1}{\sqrt3}(\ket{001}+\ket{010}+\ket{100})_{ABC}.
\end{equation}
We looked for the behaviour of $\avW$ against the angular direction of the great circle on the single-qubit Bloch sphere identified by the angles $(\theta,\phi)$. The results of our analysis are presented in Fig.~\ref{fig:simulTripartito}, where the extra work-extraction possibilities offered by the use of a GHZ-Cluster are clearly showcased. 
The study can be extended to the case of mixed-state resources by addressing the effects that a non-unit value of $\mu$ has on the performance of the work extraction protocol. This is done in Fig.~\ref{fig:simulWernerTripartito}, where we have investigated the behavior of work against $\mu$ for the following sets of angles $(\theta,\phi)=(90^{\si{\degree}},0^{\si{\degree}}),~(0^{\si{\degree}},0^{\si{\degree}})$ and $(0^{\si{\degree}},90^{\si{\degree}})$, displaying the different performance of work extraction against different choices of great circle orientation.

We have tested these predictions in a two-photon implementation of both a three-qubit GHZ-Cluster type state and a $W$ type one. Such states are generated by adding a path-encoded qubit to the two-qubit setup shown in Fig.~\ref{fig:exp_setup}. This qubit is experimentally obtained by introducing a Sagnac interferometer (consisting of a 50:50 BS and three mirrors) over the path of one of the two photons. Encoding in the path degree of freedom is then obtained by considering the logical state $\ket{0}$ (clockwise circulation of a photon inside the interferometer) and $\ket{1}$ (anticlockwise circulation). The phase $\phi$ between such logical state can be tuned by tilting a thin glass plate, placed in one of the paths. This arrangement allowed us to engineer a GHZ-Cluster state of the form $(|HH0\rangle-|VV0\rangle+|HV1\rangle+|VH1\rangle)_{ABC}/2$ by introducing a HWP at 0${\si{\degree}}$ on the clockwise path $\ket{0}$ and a HWP at 45${\si{\degree}}$ on the anticlockwise one [cf. Fig.~\ref{fig:set-upGHZ} {\bf (a)}]. On the other hand, a $W$-type resource state of the form $\ket{W}=(|HH1\rangle+|HV0\rangle+|VH0\rangle)/\sqrt{3}$ can be easily engineered by modifying the configuration for the GHZ-Cluster state. After changing the label of the two paths, a PBS can be added in the setup to perform a  polarization-into-path mapping, and thus eliminate the contribution from state $\ket{VV1}$ [cf. Fig.~\ref{fig:set-upGHZ} {\bf (b)}].

Path-qubit analysis is performed by either selecting one of the two paths, which implements a projection in the computational basis, or selecting the proper phase shift $\phi$ between the two modes, thus enabling the projection on the diagonal basis ($\phi=0,\pi$) and the circular basis ($\phi=\pi/2,-\pi/2$).

\begin{table}[t!]
  \begin{center}
    \begin{tabular}{ccc}
\hline
\hline
      $\avW^W_1$ & $\avW^W_2$ & $\avW^W_3$\\
      \hline
      $0.4715 \pm 0.0036$ & $0.6625 \pm 0.0050$ & $0.5966 \pm 0.0050$\\
      $0.4715 \pm 0.0036$ & $0.3205 \pm 0.0023$ & $0.3282 \pm 0.0027$\\
      \hline
      \hline
    \end{tabular}
  \end{center}
 \caption{Extractable work in bits for the three permutations of roles in the protocol for work extraction-based inseparability evaluated using a $GHZ$-type resource. The three columns are for the value of the extractable work achieved from the path-encoded qubit, the first polarization-encoded qubit, and the second polarization-encoded one, respectively. The first row refers to a tripartite entangled $W$ state very close to an ideal $\ket{W}$ state: The fidelity between the experimental resource and the ideal state $(\ket{HH1}+\ket{HV0}+\ket{VH0})/\sqrt3$ is $F=0.823\pm0.006$. 
The second row refers to the same $W$ configuration where, however, any coherence between path-encoded and polarization-encoded qubits is removed. Correspondingly, $\avW^W_2$ and $\avW^W_3$ are below the threshold for separability, showing that in the the corresponding configuration we only have bipartite entanglement.  }
     \label{tab:table3}
\end{table}



In all cases, there is a significant violation of the separability bound. Moreover, the symmetry  among the results that we have achieved capture unambiguously the GHZ-Cluster character of the experimental resource. In order to show this feature more clearly, we have repeated the experiment when complete decoherence is introduced in one of the qubits of the system, in an attempt to render our resource state bi-separable and wash out its genuine tripartite entanglement. When full decoherence is introduced in the path-encoded qubit by inserting a thick glass plate on one of the paths, allowing to disrupt coherence between the two modes (independent of polarization), we get $\avW_1=0.799\pm0.006$, which is very close to the corresponding value in Table~\ref{tab:table1}. However, decoherence lessens the values of the other two witnesses, to values that are very close to the separability threshold in Eq.~\eqref{ghz}, demonstrating the disappearance of tripartite entanglement.

A similar analysis has been conducted using the experimental $W$-type resource, whose corresponding experimental results are reported in Table~\ref{tab:table3}. As done for the GHZ-Cluster state, we have assessed both the case of a close-to-ideal resource and that of a strongly decohered one close to the separability threshold. The diversity (in terms of entanglement-sharing structure) of such resources is fully captured by the extractable work.

A comparison with non-locality-based criteria similar to the one reported for the study of bipartite entanglement can be made in this tripartite context. Here, rather than concentrating on two-qubit correlations, we shall investigate their genuinely multipartite nature, whose characterisation is clearly a much more demanding problem to tackle. A powerful tool in this respect is embodied by genuinely tripartite versions of a Bell inequality, such as the inequality proposed by Svetlichny~\cite{svetlichny}, whose violation witnesses the occurrence of multipartite non-local correlations. Such inequality has been successfully experimentally violated using a photonic GHZ state~\cite{lavoie} and used to ascertain the properties of the ground state of a many-body system in a photonics quantum simulator~\cite{adeline}.

In the Svetlichny game, the daemons locally rotate their respective qubit by an angle $\alpha_j$ ($j=A,B,C$) through the operator $\hat{\cal R}(\alpha_j)=\cos\alpha_j\sigma_z+\sin\alpha_j\sigma_x$ and project it over the basis of $\sigma_z$. If state $\ket{0}$ ($\ket{1}$) is found, they attach a dicothomic variable the value $+1$ ($-1$). This allows them to build the the statistical correlation function for local spin measurements $E(\alpha_A,\alpha_B,\alpha_C)$ and, in turn, construct the Mermin-Ardehali-Belisnskii-Klyshko function~\cite{mermin}
\begin{equation}
\begin{aligned}
M_3&=E(\alpha_A,\alpha_B,\alpha'_C)+E(\alpha_A,\alpha'_B,\alpha_C)\\
&+E(\alpha'_A,\alpha_B,\alpha_C)-E(\alpha'_A,\alpha'_B,\alpha'_C).
\end{aligned}
\end{equation}
The Svetlichny function is thus ${\cal S}_3=|M_3+M'_3|$, where $M'_3$ is the same as $M_3$ with $\alpha_j\leftrightarrow\alpha'_j$. Any biseparable state satisfies the inequality $|{\cal S}_3|\le 4$. Tripartite entangles states violate the inequality up to the maximum value of $4\sqrt2=5.65685$, which is obtained using GHZ-like states. In fact, state $\ket{GHZc}$ allows to achieve such a maximum violation for $(\alpha_A,\theta_B,\theta_C)=(3\pi/8,-\pi/4,0)$ and $(\alpha'_A,\theta'_B,\theta'_C)=(\pi/8,0,\pi/4)$. We have used the experimental quantum state tomographies of the GHZ-Cluster and $W$ resource states to evaluate the Svetlichny function. After a global optimisation over all the angles involved, we have found ${\cal S}^{GHZc}_3=4.83$ and ${\cal S}^{W}_3=3.39$. This shows that the experimental GHZ-Cluster state is consistently found to be entangled in a genuinely tripartite sense by both the extractable work-based criterion and the Svetlichny one. On the other hand, the non locality-based entanglement criterion fails to detect the tripartite entangled nature of the experimental $W$ resource, which is instead well captured by the sensitivity exhibited by the extractable work.


\section{Conclusions}
\label{final}

We have demonstrated experimentally a fundamental result of information thermodynamics, showing that entangled working media are able to provide a significant surplus in the amount of work that can be extracted through a communication assisted game fundamentally based on Maxwell's daemon. Such protocols, in turn, represent  viable tests for the inseparability of a given state resource that, as demonstrated in our experimental endeavors, are both practical and fundamentally interesting in light of their distinct nature from, say, non-locality based entanglement witnesses. This work has significant implications for both technological and conceptual aspects. Concerning the technological advances that it entails, we contribute to the well sought-after and ongoing efforts aimed at setting up a photonics-based platform for information thermodynamics. Such platform has the capability of preparing, controlling and measuring, with high fidelity and a reduced experimental complexity, states of a multipartite working medium. Furthermore, general evolutions can be easily implemented or simulated, thus paving the way to the implementation of a fully fledged experimental assessment of non-equilibrium dynamics. The latter has so far been limited to either classical systems or nuclear magnetic resonance setups, where any quantum feature is weakened by the typically strong environmental actions. As for the fundamental aspects, we have highlighted how non-classical correlations within the working medium should be interpreted as a resource for the performance of thermodynamic processes, not differently from quantities with a counterpart in classical thermodynamics. This points to an interesting direction for understanding the emergence of ordinary world from its quantum microscopic constituents.


\acknowledgments

We are grateful to M. Vidrighin and V. Vedral for insightful feedback on this manuscript. This work was partially supported by the project FP7-ICT-2011-9-600838 (QWAD Quantum Waveguides Application and Development; www.qwad-project.eu). MB is supported by a Rita Levi-Montalcini fellowship of MIUR. MP acknowledges financial support from the EU FP7 Collaborative Project TherMiQ, the John Templeton Foundation (grant number 43467), the Julian Schwinger Foundation (grant number JSF-14-7-0000), and the UK EPSRC (grant number EP/M003019/1).

\end{document}